\documentclass[preprint,prl,showpacs,superscriptaddress,
draft,amsmath,amssymb]{revtex4}

\usepackage{graphicx}
\usepackage{dcolumn}
\usepackage{bm}

\makeatletter

\def\compoundrel#1\over#2{\mathpalette\compoundreL{{#1}\over{#2}}}
\def\compoundreL#1#2{\compoundREL#1#2}
\def\compoundREL#1#2\over#3{\mathrel
    {\vcenter{\hbox{$\m@th\buildrel{#1#2}\over{#1#3}$}}}}
\makeatother

\begin{document}

\title{Observation of superconducting gap in boron-doped 
diamond by laser-excited photoemission spectroscopy}

\author{K. \surname{Ishizaka}}
\affiliation{Institute for Solid State Physics, University of Tokyo,
 Kashiwa, Chiba 277-8581, Japan}

\author{R. \surname{Eguchi}}
\affiliation{The Institute of Physical and Chemical Research (RIKEN),
Sayo-gun, Hyogo 679-5143, Japan}

\author{S. \surname{Tsuda}}
\affiliation{Institute for Solid State Physics, University of Tokyo,
 Kashiwa, Chiba 277-8581, Japan}

\author{T. \surname{Yokoya}}
\affiliation{The Graduate School of Natural Science and Technology, 
Okayama University, Okayama 700-8530, Japan}

\author{T. \surname{Kiss}}
\affiliation{The Institute of Physical and Chemical Research (RIKEN),
Wako, Saitama 351-0198, Japan}

\author{T. \surname{Shimojima}}
\affiliation{Institute for Solid State Physics, University of Tokyo,
 Kashiwa, Chiba 277-8581, Japan}

\author{T. Togashi}
\affiliation{The Institute of Physical and Chemical Research (RIKEN),
Sayo-gun, Hyogo 679-5143, Japan}

\author{S. Watanabe}
\affiliation{Institute for Solid State Physics, University of Tokyo,
 Kashiwa, Chiba 277-8581, Japan}

\author{C.-T. Chen}
\affiliation{Beijing Center for Crystal R\&D, Chinese Academy of Science,
Zhongguancun, Beijing 100080, China}

\author{C. Q. \surname{Zhang}}
\affiliation{Beijing Center for Crystal R\&D, Chinese Academy of Science,
Zhongguancun, Beijing 100080, China}

\author{Y. \surname{Takano}}
\affiliation{National Institute for Material Science, 
Tsukuba, Ibaraki 305-0047, Japan}

\author{M. \surname{Nagao}}
\affiliation{National Institute for Material Science, 
Tsukuba, Ibaraki 305-0047, Japan}

\author{I. \surname{Sakaguchi}}
\affiliation{National Institute for Material Science, 
Tsukuba, Ibaraki 305-0047, Japan}

\author{T. \surname{Takenouchi}}
\affiliation{School of Science and Engineering, Waseda University,
Shinjuku, 
Tokyo 169-8555, Japan}

\author{H. \surname{Kawarada}}
\affiliation{School of Science and Engineering, Waseda University,
Shinjuku, 
Tokyo 169-8555, Japan}

\author{S. \surname{Shin}}
\affiliation{Institute for Solid State Physics, University of Tokyo,
 Kashiwa, Chiba 277-8581, Japan}
\affiliation{The Institute of Physical and Chemical Research (RIKEN),
Sayo-gun, Hyogo 679-5143, Japan}

\begin{abstract}

We have investigated the low energy electronic state of a
boron-doped diamond thin film by the ultrahigh resolution
laser-excited photoemission spectroscopy. 
We observed a clear shift of the leading edge  
below 11 K indicative of a superconducting gap opening
($\Delta \sim 0.78$ meV at 4.5 K). 
The gap feature is significantly broad and
the well-defined quasiparticle 
peak is not recognizable even at the lowest temperature of 
measurement 4.5 K.
We discuss our result in terms of possible disorder 
effect on superconductivity in this system.

\end{abstract}

\maketitle


Diamond has been well studied through the 
progress of solid state physics as one of the simplest 
$sp^3$ covalent system.
Although it is a typical band insulator with a gap of 5.5 eV, 
its high thermal conductivity owing to the 
high Debye temperature is well known also from the
application point of view.
When doped with boron (B), which plays a role of an accepter 
with the activation energy of 0.37 eV, the diamond behaves as a
well organized $p$-type semiconductor.
On increasing the B concentration ($n_{\rm B}$), the system goes across 
the insulator-metal (Mott) transition at 
$n_{\rm B}^{\rm MI} \sim 2\times 10^{20}$ 
cm$^{-3}$.
It is recently that a striking discovery of the superconductivity
($T_{\rm c}\sim 2 $ K) was reported 
in B-concentration above $n_{\rm B}^{\rm MI}$ \cite{ekimov04}.
Shortly after, by virtue of advanced crystal growth technique 
such as the chemical vapor deposition (CVD) method, 
the sample quality and its $T_{\rm c}$ have become
higher \cite{takano04,takanoDRM,bustarret04,umezawa05}.
Such progress has made possible the
experimental studies in a wide variety, 
and opens a new field for device applications using 
diamond related materials.

From the theoretical side, several scenarios have been 
proposed. 
The band calculations with the virtual crystal approximation (VCA) 
\cite{boeri04,lee04}
predict that the top of the host diamond valence band at $\Gamma $-point 
is shifted 
above the Fermi level in the metallic region 
({\it e.g.} 0.61 eV shift for 2.5\% hole doping in \cite{lee04}). 
The holes at the small pocket around $\Gamma $ strongly 
couple to the optical phonon with $\sim 160 $ meV, 
which gives rise to the BCS-type superconductivity. 
Actually, the rigid shift of the valence band and the 
phonon softening at $\Gamma$-point are recently confirmed by the
angle-resolved PES \cite{yokoya05} and 
inelastic x-ray \cite{hoesch05} measurements respectively, 
which shows that the above models are valid at least qualitatively.
The supercell calculation
points out a certain importance of the boron circumstance in addition, 
which has a large local density-of-states (DOS) at $E_{\rm F}$ 
(though smaller than the total DOS of carbon)
and modifies the host diamond band by its randomness
\cite{blase04,xiang04}.
They also show that the local B-C vibration mode will significantly 
contribute to the 
electron-phonon coupling constant $\lambda $.
On the contrary, there is a theoretical report which is 
entirely based upon 
the half-filled narrow
boron impurity band as schematically realized in the Mott 
transition picture \cite{baskaran04}. 
This scenario will produce a dirty superconducting state with 
an extended $s$-wave gap by 
an exotic pairing interaction, resonating-valence-bond (RVB)
mechanism, which manifests itself in a strong correlation 
regime.

In this study, we investigated the low-energy electronic structure 
of the B-doped diamond to clarify the characteristics of the 
superconductivity realized in this doped semiconductor system.
 By using the ultrahigh resolution 
laser-excited photoemission spectroscopy, we succeeded in 
the observation of the superconducting gap 
evolving below 11 K.
We discuss our result in the light of boron-induced 
disorder effect.

PES measurements were performed using a system constructed with 
the Scienta R4000 electron analyzer and an ultraviolet ($h\nu =6.994$ eV) 
laser for the incident light \cite{kiss05}. 
The escape depth of the photoelectron in this energy region 
({\it i.e.} kinetic energy of $2 \sim 3$ eV) attains a large value of
$\sim 100$ \AA, 
\cite{seah79} which enables us the bulk-sensitive measurements.
The temperature was precisely controlled
down to 4.5 K using a flow-type He liquid refrigerator. 
The base pressure of the chamber was below $\sim 5\times 10^{-11}$ Torr
throughout all the measurements.
We annealed the samples before the measurement at 400 \(^\circ\)C
in the vacuum of $< 10^{-7}$ Torr.
This procedure increased the 
photoelectron intensity near the Fermi level ($E_{\rm F}$) 
for about a factor of 10 times, while not changing the 
spectral shape.
Since we could not catch clear evidence of angle-resolved 
($k$-dispersed) signal,
all measurements were performed with an angle-integration mode.
The energy resolution was
$\Delta E = $ $0.7 $ meV.
The Fermi level ($E_{\rm F}$) of the sample was referred to that of 
the Au film 
evaporated on the sample substrate.

The B-doped diamond (BDD) sample we used for our measurement
is obtained by a microwave plasma-assisted 
chemical vapor deposition (CVD) method
as described elsewhere \cite{takano04,umezawa05}.
It was grown homoepitaxially on an undoped (1 1 1) oriented 
diamond synthesized in high pressure and high temperature
\cite{umezawa05}. 
The BDD thus grown is single crystalline.
The boron concentration determined by the 
secondary ion mass spectroscopy method is 
$n_{\rm B}=8.4 \times 10^{21}$ ${\rm cm}^{-3}$, 
which corresponds to B/C ratio of 5\%.
The carrier density estimated from the 
Hall coefficient is $n_{\rm H}=1.3 \times 10^{22}$ ${\rm cm}^{-3}$
\cite{umezawa05}.
We show in Fig. \ref{fig1} the resistivity and 
magnetization curves for the BDD sample.
The superconducting transition temperature of this sample
determined from the Meissner response is 
$T_{\rm c}^{\rm m}= 6.6 $ K as indicated in 
the inset of Fig. \ref{fig1}(b).
The diamagnetic response keeps increasing on lowering the 
temperature down to 2 K without any sign of saturation,
reflecting the disordered nature of the superconductivity.
The resistivity, on the other hand, 
steeply starts decreasing at 8.2 K
and shows the zero-resistivity at around 7 K.
There is another characteristic temperature 
$T_0 \sim 11$ K where the resistivity starts to deviate from 
that of the weakly insulating normal state under the magnetic 
field, which will be discussed later.
The normal-state resistivity is almost temperature-independent
with the values of $\rho ({\rm 300K})=0.59$ m$\Omega $ (not shown) 
and $\rho ({\rm 10K}) =0.68 $ m$\Omega$, indicative of the
fairly localized carrier dynamics in this system.
Here we estimate the carrier mean free path from 
$l=\frac{\hbar k_{\rm F} \tau}{m^*}=
\frac{\hbar k_{\rm F}}{ne^2\rho}$ and
$k_{\rm F}=(3\pi ^2 n)^{1/3}$ in a simple free-electron picture.
Substituting $\rho ({\rm 10K})$ and the above mentioned 
$n_{\rm H}$ for the carrier density $n$, 
we obtain $l=3.4$ \AA.
It is nearly equivalent with the 
cubic lattice parameter $a=3.57$ \AA \cite{umezawa05}
indicating that this system is fairly close to the
Mott-Ioffe-Regel limit \cite{takagi92}.
We also note that this value coincides with that
previously reported by Sidorov {\it et al.} in a 
lower-doped single crystalline 
sample ($n_{\rm H} \sim 1.8 \times 10^{21}$ cm$^{-3}$, 
$\rho ({\rm 10K}) \sim 2.5 $ m$\Omega$cm) \cite{sidorov05}.
Since $k_{\rm F}$ is greater in our sample with higher 
$n$, the carrier mobility $\mu = \frac{e \tau}{m^*}$ must be  
lower instead in order to obtain the same 
$l$, where $\tau  $ and 
$m^*$ represent the lifetime and the effective mass of the carriers.
On the other hand, the coherence length estimated 
from the upper critical field $H_{\rm c2}$ is 
$\xi \sim 60$ \AA $> l$ \cite{umezawa05}, 
which classifies this BDD sample 
as a dirty superconductor.


Figure \ref{fig2} shows the photoelectron spectrum (PES) 
of the superconducting state 
observed in the BDD sample.
With decreasing temperature from 15.5 K, a slight shift of the 
spectral edge can be observed below 9.5 K
(see the enlarged plot in Fig. \ref{fig2}(b)).
At the same time, the spectral shape becomes
slightly convex downward at $E_{\rm F}$.
Such temperature dependence is indicative of a 
gap opening at $E_{\rm F}$, which is a clear evidence of 
the mass superconductivity in this sample. 
Even at the lowest temperature (4.5 K), however, 
a considerable amount of DOS at $E_{\rm F}$ remains 
and a clear quasiparticle peak cannot be observed.
We tried to estimate the gap value at 4.5 K by fitting the 
PES using the Dynes function 
\cite{dynes78} for the DOS represented with 
the isotropic ($s$-wave) superconducting gap 
and the phenomenological 
broadening parameters $\Delta $ and $\Gamma $, as
$$D(E_{\rm B}, \Delta, \Gamma)
={\rm Re}  
\frac{E_{\rm B}-i\Gamma}{\sqrt{\left(E_{\rm B}-i\Gamma \right)^2-\Delta^2}}
~~.$$ 
The best-fitted result is shown in Fig. \ref{fig2}(c) 
as the red curve, with the parameters $\Delta = 0.78$ meV 
and $\Gamma = 0.7$ meV. 
We note that $\Gamma $ is remarkably large,
$\Gamma \approx \Delta$, which accounts for the 
large DOS at $E_{\rm F}$ and the hardly recognizable 
quasiparticle peak.
Due to such spectral shape, 
it is impossible at present to
discuss the superconducting gap symmetry, 
whether it is  simple $s$-wave or not,
solely from our result.
Here we estimate the reduced gap 
$\Delta (0)/k_{\rm B} T_{\rm c}$ from our result at the 
lowest temperature, ie.  
$\Delta ({\rm 4.5K})=0.78$ meV,
by simply assuming the BCS-like temperature dependence of 
$\Delta (T)$.
If we take $T_{\rm c}=T_{\rm c}^{\rm m}=6.6$ K, 
$\Delta (0)/k_{\rm B} T_{\rm c}= 1.78$ is obtained with 
$\Delta (0)=1.01$ meV.
It is close to a typical value for weakly-coupled BCS 
superconductors.
This result thus seems to suggest that $\Delta $ we observed 
 is well related in a weak-coupling regime 
with $T_{\rm c}^{\rm m}$ where the 
superconductivity shows up in the volume-sensitive magnetization.

To confirm the evolution of the gap structure itself, 
we show in Fig. \ref{fig3} the 
PES symmetrized at $E_{\rm F}$ to remove the 
spectral cutoff by the Fermi-Dirac distribution.
It is now clearly recognized that the intensity at $E_{\rm F}$ 
starts to decrease and gradually forms a gap below 9.5 K. 
Also in this symmetrized PES which represents the DOS, 
we cannot discern a well-defined quasiparticle peak.
This result is in a striking contrast to 
that of scanning tunneling spectroscopy (STS) study
on a (1 0 0) CVD thin film with 
$T_{\rm c}=1.9 $ K and $n_{\rm B}=1.9\times 10^{21}$ 
cm$^{-3}$ \cite{sacepe05}, 
where a superconducting gap spectrum with well defined 
quasiparticle peaks 
($\Delta =0.285$ meV and $\Gamma = 255$ mK $=0.022$ meV at $T= 70 $ mK)
is observed. 
Furthermore, its temperature dependence is well in accord with 
the weak-coupling BCS gap function with
$\Delta (0) = 1.74 k_{\rm B} T_{\rm c}$. 
On the other hand, another very recent STS result on a 
(1 1 1) CVD thin film with $n_{\rm B}=6\times 10^{21}$ cm$^{-3}$
and $T_{\rm c}^{\rm m}=5.4$ K
shows a broad superconducting 
gap spectrum fairly similar to ours with 
$\Delta =0.87 $ meV and $\Gamma =0.38 $ meV at 0.47 K
\cite{nishizaki06}.
Both measurements report that the STS spectra reflecting 
the local DOS show very little 
dependence on the location at the sample surfaces, 
which should rule out the possibility of mesoscopic (nanoscale)
modification of the superconducting state in these samples.
The discrepancy among the results on 
(1 0 0) and (1 1 1) samples 
may be explained by the difference in $n_{\rm B}$.
We note that on increasing $n_{\rm B}$, the mobility 
tends to get lower reflecting the disorder induced by 
random boron doping.
Recent ARPES result also shows that the lifetime of the 
carriers estimated from the spectral linewidth becomes 
shorter in the sample with higher $n_{\rm B}$ \cite{yokoya05}.
They indicate the possible dirtiness of 
the electronic structure in high-$n_{\rm B}$ samples, 
which gives rise to the superconducting gap 
with large $\Gamma $ as observed in (1 1 1) samples.
Another possibility is the extrinsic 
sample inhomogeneity which is 
known to be greater in (1 1 1) samples. 
It is actually reported that there are two 
structural phases with slightly 
different lattice constants in (1 1 1) samples when 
$n_{\rm B}$ and $T_{\rm c}$ become high 
($T_{\rm c}^{\rm m} > 6$ K) \cite{takenouchi06}. 
Since the lattice constant of a heavily B-doped diamond 
is greater than that of undoped 
ones by about 0.5\%, 
there is a tendency toward a uniaxially expanded phase 
near the substrate and 
an isotropically relaxed phase for the rest. 
Though exactly how these two phases coexist in BDD thin films is 
yet under investigation, 
such local inhomogeneity may account for the lousy superconducting 
gap structure in our PES result. 

Now we discuss the temperature dependence of the superconducting 
gap in this sample. 
Since it is very difficult to get accurate gap values by 
fitting these rather featureless spectra 
for all temperatures, we simply estimated
the energy shift of the leading edge in PES (Fig. \ref{fig2}) 
instead.
Its temperature dependence in Fig. \ref{fig4} 
rather resembles that of the Meissner response which
monotonically keeps growing on decreasing temperature, 
as shown in Fig. \ref{fig1}(b).
The onset temperature of gap evolution 
 at around $T_0 \sim 11$ K, 
however, is apparently 
higher than that of the magnetization, $T_{\rm c}^{\rm m}=6.6 $ K. 
Looking back on the resistivity curve in Fig. \ref{fig1}(a), 
we find a characteristic behavior at around $T_0$.
At $T_0$, the resistivity starts to deviate from the
weakly insulating normal state which is apparent under 
a magnetic field.
Such a deviation may be reflecting the inhomogeneous conductivity 
near $T_{\rm c}$, which attributes $T_0$ to the 
onset of the local superconducting transition and 
$T_{\rm c}^{\rm off} \sim 7$ K to the temperature of  
bulk supercurrent percolation.
If this is the case, our PES result is suggestive of its 
high sensitivity that probes the local superconducting 
state with high $T_{\rm c}$.
At the same time, it shows the potential for development of a
diamond superconductor with $T_{\rm c} \sim 11$ K 
by aggregating such local ``high-$T_{\rm c}$" segments.

In conclusion, we have performed an ultrahigh resolution 
photoemission spectroscopy measurement to elucidate the 
near-$E_{\rm F}$ electronic structure of the 
superconducting boron-doped diamond
with $T_{\rm c}^{\rm m} =6.6 $ K.
We observed the 
gradual formation of the superconducting gap below 
$T_{0} \sim 11$ K, the temperature where the 
deviation of the resistivity from the normal state is observed.
The dominant size of the gap, though under an uncertainty due to the 
lack of quasiparticle peak,  
is estimated to be about $\Delta =0.78$ meV at 4.5 K. 
Its broad spectral shape and the diffusive temperature 
dependence indicate that the superconductivity in this system 
is strongly affected by randomness and/or inhomogeneity 
introduced by boron doping.
Further precise investigation is desired using samples with 
the least extrinsic nonstoichiometry.

\newpage

\begin{figure}[htbp!]
\caption{
(color online). Temperature dependence of the 
resistivity (a) and the magnetization (b) in 
B-doped diamond.
Insets show the magnification of the 
temperature regions near the superconducting  
transition.
$T_0$ shows the temperature where the resistivity starts 
to deviate from the normal state behavior, whereas 
$T_{\rm c}^{\rm m}$ represents the onset temperature of the 
Meissner response.
\label{fig1}
}
\end{figure}

\begin{figure}[htbp!]
\caption{
(color). Temperature dependence of the 
ultrahigh resolution photoemission spectrum in the 
B-doped diamond (a). 
Enlarged plot around $E_{\rm F}$ is shown in (b). 
(c) shows the PES at 4.5 K indicated together 
with the fitting result.
\label{fig2}
}
\end{figure}

\begin{figure}[htbp!]
\caption{
(color online). Temperature dependence of PES in B-doped diamond 
symmetrized at $E_{\rm F}$ to exclude the 
Fermi-Dirac cutoff.
\label{fig3}
}
\end{figure}

\begin{figure}[htbp!]
\caption{
(color online). Temperature dependence observed in the energy shift of the 
leading edge in PES (Fig. \ref{fig2}), which should reflect 
that of the superconducting gap. 
$T_0$ and $T_{\rm c}^{\rm m}$ indicate the characteristic 
temperature observed in resistivity and magnetization, 
respectively (see Fig. \ref{fig1}).
Broken curve is merely a guide-for-eyes.
\label{fig4}
}
\end{figure}


\begin{thebibliography}{99}

\bibitem{ekimov04} 
E. A. Ekimov, V. A. Sidorov, E. D. Bauer, N. N. Mel'nik, N. J. Curro,
J. D. Thompson, and S. M. Stishov, Nature (London) 
{\bf 428}, 542 (2004).

\bibitem{takano04} 
Y. Takano, M. Nagao, I. Sakaguchi, M. Tachiki, T. Hatano, K. Kobayashi, 
H. Umezawa, and H. Kawarada, 
Appl. Phys. Lett. {\bf 85}, 2851 (2004).


\bibitem{takanoDRM} 
Y. Takano, M. Nagao, T. Takenouchi, 
H. Umezawa, I. Sakaguchi, M. Tachiki, and H. Kawarada, 
Diam. Relat. Mater. {\bf 14}, 1936 (2005).



\bibitem{bustarret04} 
E. Bustarret, J. Kacmarcik, C. Marcenat, E. Gheeraert, C. Cytermann, 
J. Marcus, T. Klein,  
Phys. Rev. Lett. {\bf 93}, 237005 (2004).

\bibitem{umezawa05}
H. Umeazawa, T. Takenouchi, Y. Takano, K. Kobayashi, M. Nagao, 
I. Sakaguchi, A. Ishii, M. Tachiki, T. Hatano, G. Zhong, 
M. Tachiki, and H. Kawarada, 
cond-mat/0503303 (unpublished).

\bibitem{boeri04} 
L. Boeri, J. Kortus, and O. K. Andersen, 
Phys. Rev. Lett. {\bf 93}, 237002 (2004).

\bibitem{lee04} 
K. W. Lee and W. E. Pickett, 
Phys. Rev. Lett. {\bf 93}, 237003 (2004).

\bibitem{yokoya05}
T. Yokoya, T. Nakamura, T. Matsushita, T. Muro, Y. Takano, 
M. Nagao, T. Takenouchi, H. Kawarada, and T. Oguchi, 
Nature (London) {\bf 438}, 648 (2005).

\bibitem{hoesch05}
M. Hoesch, T. Fukuda, T. Takenouchi, J. P. Sutter, S. Tsutsui, 
A. Q. R. Baron, M. Nagao, Y. Takano, H. Kawarada, and J. Mizuki, 
cond-mat/0512424 (unpublished). 

\bibitem{blase04} 
X. Blase, Ch. Adessi, and D. Connetable, 
Phys. Rev. Lett. {\bf 93}, 237004 (2004).

\bibitem{xiang04} 
H. J. Xiang, Z. Li, J. Yang, J. G. Hou, and Q. Zhu, 
Phys. Rev. B {\bf 70}, 212504 (2004). 



\bibitem{baskaran04}
G. Baskaran, cond-mat/0404286 (unpublished).


\bibitem{kiss05} 
T. Kiss, F. Kanetaka, T. Yokoya, T. Shimojima, K. Kanai, S. Shin, 
Y. Onuki, T. Togashi, C. Zhang, C. T. Chen, and S. Watanabe,
Phys. Rev. Lett. {\bf 94}, 057001 (2005).


\bibitem{seah79} 
M. P. Seah and W. A. Dench,
Surf. Interface Anal. {\bf 1}, 2 (1979).

\bibitem{takagi92} 
H. Takagi, B. Batlogg, H. L. Kao, J. Kwo, R. J. Cava, 
J. J. Krajewski, and W. F. Peck Jr., 
Phys. Rev. Lett. {\bf 69}, 2975 (1992); 
N. E. Hussey, K. Takenaka, and H. Takagi, 
cond-mat/0404263.



\bibitem{sidorov05}
V. A. Sidorov, E. A. Ekimov, S. M. Stishov, E. D. Bauer, 
and J. D. Thompson, 
Phys. Rev. B {\bf 71}, 060502(R) (2005).

\bibitem{dynes78}
R. C. Dynes, V. Narayanamurti, and J. P. Garno, 
Phys. Rev. Lett. {\bf 41}, 1509 (1978).

\bibitem{sacepe05}
B. Sacepe, C. Chapelier, C. Marcenat, J. Kacmarcik, T. Klein, 
M. Bernard, and E. Bustarret, 
cond-mat/0510541 (unpublished). 

\bibitem{nishizaki06}
T. Nishizaki, Y. Takano, M. Nagao, T. Takenouchi, H. Kawarada, 
and N. Koabayashi, (unpublished).


\bibitem{takenouchi06}
T. Takenouchi, S. Tezuka, H. Ishiwata, Y. Takano, M. Nagao, 
M. Tachiki, I. Sakaguchi, and H. Kawarada, (unpublished).



\end{thebibliography}
\end{document}